\pdfoutput=1
\documentclass{ms}

\usepackage{amsmath}

\shorttitle{High-Latitude Rotation Rate}
\shortauthors{Sheeley}

\graphicspath{{./}}

\begin{document}

\title{Using Polar Faculae to Determine the Sun's High-Latitude Rotation Rate. I. Techniques and Initial Measurements}

\author[0000-0002-6612-3498]{Neil R. Sheeley, Jr.}
\affiliation{Visiting Research Scientist\\
Lunar and Planetary Laboratory, University of Arizona \\
Tucson, AZ 85721, USA} 

\begin{abstract}
This paper describes a new way of determining the high-latitude solar rotation rate statistically
from simultaneous observations of many polar faculae.  In this experiment, I extracted 
frames from a movie made previously from flat-fielded images obtained in the 6767 {\AA} continuum
during February 1997-1998 \citep{SHEWAR_2006}
and used those frames to construct space-time maps from high-latitude slices of the favorably
oriented south polar cap.  These maps show an array of slanted tracks whose average slope
indicates the east-west speed of faculae at that latitude, ${\lambda}_{s}$.  When the slopes are measured and plotted as a function of latitude, they show relatively little scatter
${\sim}$0.01-02 km s$^{-1}$ from a straight line whose zero-speed extension passes through the Sun's south pole.  This means that the speed, $v({\lambda}_{s})$, and the latitudinal
radius, R$_{\odot}$$\cos{\lambda}_{s}$, approach 0 at
the same rate, so that their ratio gives a nearly constant synodic rotation rate
${\sim}$8.6$^{\circ}$ day$^{-1}$ surrounding the Sun's south pole.  A few measurements of the unfavorably oriented north polar cap are consistent with these measurements near the south pole.
\end{abstract}

\keywords{Solar faculae (1494)--- Solar rotation (1524)---Solar magnetic fields (1503)---Solar cycle (1487)}

\section{Introduction} \label{sec:intro}
The idea for this study occurred on 26 June 2024 when I learned about the COFFIES
(Consequences of Fields and Flows in the Interior and Exterior of the Sun) NASA-funded virtual
workshop on `The Science of the Poles', organized by Lisa Upton, Shea Hess Webber, and
Todd Hoeksema.  In that workshop, my work on polar faculae
 \citep{SHEfac_1964,SHE_1965,SHEfac_1976,SHEfac_1991,SHEfac_2008,SHEfac_2012}
was mentioned.  Those studies were about counting polar faculae on white-light images
obtained at the Mount Wilson Observatory since 1905 and plotting their evolution over
many sunspot cycles.  However, in 2006, Harry Warren and I conducted a different study of
faculae using images obtained in the 6767 {\AA} continuum during 1996-2005 with the Michelson Doppler Interferometer (MDI) on the \textit{Solar and Heliospheric Observatory} (SOHO) \citep{SHEWAR_2006}, and I wondered if that study might be relevant to the polar studies
of the COFFIES group too.  

Our main objective in that 2006 study was to make Carrington maps that showed
the distribution of faculae over the full Sun, similar to Carrington maps of photospheric
magnetic fields and the associated Ca II K-line emission seen in Mount Wilson spectroheliograms
\citep{SHECOOPA_2011}.  However, in that 2006 paper, we included a figure with an online link to a
time-lapse movie of the 6767 {\AA} disk.  The movie consisted of flat-fielded
images\footnote{Jeneen Sommers (Stanford University) had provided those images in a
flat-fielded form that showed the faculae without the presence of limb darkening.}, obtained during February 7-21 of each year when the solar $B_{0}$ angle was centered around -6.8$^{\circ}$ and
favored a view of the Sun's south pole.  By running the movie at high speed, the background noise was
effectively reduced so that the polar faculae were visible as a `cloud of bright points' waxing
and waning with time as the sunspot cycle advanced.

However, it was also easy to see individual faculae moving east-to-west with the solar rotation.  I wondered if it would be possible to capture this motion in a space-time map, obtained by placing
the slit along a chord in the south polar cap.  I wanted to keep the project simple and just
measure the speed near the central meridian.  I did not want to worry about problems that would
arise toward the ends of the slit where the curved latitude lines would bend off a straight slit
and where the curvature of the solar surface would reduce the `sky plane' speed of faculae
(which would happen even for the straight latitude contours that occur when $B_{0}=0$). 

To accomplish this feat, I enlisted the help of ChatGPT, first to extract the frames from the movie
and then to cut out narrow east-west oriented strips from these images and place them in chronological order.  I extracted the images and stored them in a directory manually, which took several hours
because I wanted to use all 106 of the images during the February 7-21 intervals of 1997 and 1998.
In those years, the polar fields were relatively strong compared to the polar fields around sunspot
minima in 2009 and 2020, so that may have been our last opportunity to track the faculae when
they were still plentiful.  Making the space-time maps was much faster than collecting the frames.
It only took a few iterations with ChatGPT to obtain a working program that showed slanted tracks.
The sight of those tracks was the first `ah-ha' moment of this study.  It was clear that the procedure would work and that success was only a matter of time and `fine tuning' the program.  (The
second `ah-ha' moment
occurred later when I discovered that the plotted data points were well fit by a straight line whose
zero-speed extension passed through the Sun's south pole.  Consequently, both the linear speed of
the faculae and the radius of their latitude contour would approach zero at the same rate and
give a nearly constant synodic rotation speed in the vicinity of the pole.)

This paper is organized as follows:
In the next section, I will describe the measurement procedure in detail and show a sample of
the resulting space-time maps, obtained with different slits and at different latitudes.  A few
measurements were obtained in the northern hemisphere despite its unfavorable view from SOHO
(located at the L1 point on the Earth-Sun line) in February.  Then, in the Results section, I show
the speed obtained from the space-time maps and plotted versus the measurement latitudes.
Finally, in the last section, I summarize these results and compare them with ground-based measurements obtained by tracking magnetic flux elements with the Mount Wilson magnetograph
\citep{SNOD_1983} and by tracking individual polar faculae when they were especially plentiful
during the years 1952-1954 \citep{ROLF_1954} and 1951-1954 \citep{WALD_1955}.
  
\section{Measurement Techniques}\label{sec:analysis_techniques}
\subsection{The measurement process}
Figure~1 is a sketch of the south polar cap where most of the measurements were made.  The red
lines indicate 60$^{\circ}$, 70$^{\circ}$, and 80$^{\circ}$ latitude contours on 14 February when
the solar $B_{0}$ angle is -6.8$^{\circ}$, and the thin horizontal lines indicate the upper and lower
edges of a 7-pixel (20 Mm) slit, in this case centered on the 70$^{\circ}$ contour.  Note that
\begin{figure}[h!]
%\plotone{fig01.pdf}
 \centerline{
 \fbox{\includegraphics[bb=65 290 550 500,clip,width=0.95\textwidth]
 {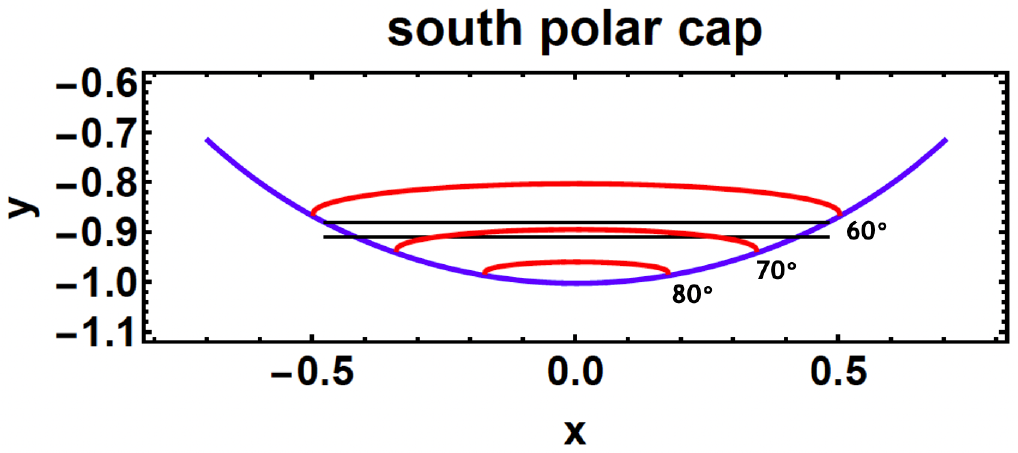}}}
\caption{Drawing of the south polar cap around 14 February when $B_{0}=-6.8^{\circ}$
and the south pole is tipped toward Earth (and SOHO).  The solar limb is drawn in blue;
 latitude contours at 60$^{\circ}$, 70$^{\circ}$, and 80$^{\circ}$ are indicted in red; and
 the upper and lower edges of a sample 7-pixel wide slit are indicated by horizontal black lines.
 The rectangular coordinate axes, x and y, are expressed in units of solar radii.}
\label{fig:fig1}
\end{figure}
\noindent
the upper edge of the slit ends at the solar limb (indicated by the solid blue curve), whereas
the lower edge crosses the limb into the sky background.  As the reader will see in the next
subsection, this gives the resulting space-time map jagged upper and lower
edges.  Finally, in reference to Figure~1, the horizontal and vertical dimensions are $x$ and
$y$ expressed in solar radii with the disk center at (0, 0) and the southernmost limb at the
point (0, -1).

However, rectangular distances on the solar image are expressed in pixels with (256, 502)
at the southernmost limb and (256,10) at the northernmost limb, and a solar radius of 246
pixels.  Consequently, I derived the latitudes of the upper and lower edges of the slit
from $y$-coordinates measured in pixels and I calculated the chord lengths from
$x$-coordinates also measured in pixels.  In particular, the latitudes were obtained using
the relation
\begin{equation}
1-\frac{(502-yp_{i})}{246}~=~-y_{i}~=~\sin({\lambda}^{(i)}_{s}-6.8),
\end{equation}
where ${\lambda}^{(i)}_{s}$ refers to the south latitude of the upper ($i=1$) or lower ($i=2$)
edge of the slit in degrees, $y_{i}$ is the $y$-value of that edge of the slit in solar radii,
$yp_{i}$ is the location of the corresponding edge in pixels,  502 is the $y$-value of the southernmost limb in pixels, and 246 is the radius of the solar image in pixels.  Finally, I calculated the average,  $<{\lambda}_{s}>$, of these two latitudes and used this average as the effective latitude in plots of the speed, 
$v({\lambda}_{s})$.

I computed the
speed by determining how long it would take a given linear track to extend from the row of
upper peaks at the east limb to the row of upper peaks at the west limb and then used the relation
\begin{equation}
v~=~\frac{{\Delta}x}{{\Delta}t},
\end{equation}  
where ${\Delta}x$ is the maximum chord length in solar radii given by
${\Delta}x = (x_{2}-x_{1})/246$, where $x_{1}$ and $x_{2}$ are the $x$-coordinates of the left and
right ends of the upper edge of the slit in pixels, and ${\Delta}t$, is the corresponding time difference, obtained from the space-time map.

In practice, this time measurement was done by using a ruler to fit the array of approximately
parallel tracks with a straight line and then measuring the horizontal distance that was cut off
between the ends of that line in millimeters.  This `time' was calibrated in days by comparing the total duration of the map in millimeters with its corresponding interval in days.  This interval was 30 days for the 106 frames used in the two intervals 7-21 February 1997 and 7-21 February 1998 because the end dates of 7 February and 21 February were used in both 1997 and 1998.   It was important to use the correct time difference (30 days rather than 28 days without the end dates) because its value directly affects the value of the final synodic rotation rate.

An interesting aspect of this unusual movie is that some of the frames in 1997 were
not displayed with the same cadence as others.  As the mission evolved and more time became
available, the cadence increased from a few frames per day in 1997 to a precise routine of 4 frames
per day in 1998.  In 1997, there were 6 days with 2 frames, 4  days with 3 frames, and 5 days with
4 frames for an average of almost exactly 3 frames per day.  So our 30-day interval consisted of
15 days with frames at an average rate of 3 frames per day followed by 15 days with frames at a
precise cadence of 4 frames per day.  This may have caused some of the tracks to wiggle a little,
but it should not have affected the overall calibration of 30 days.
   
\subsection{The space-time maps}
Figure~2 shows space-time maps obtained at an average latitude of approximately 76$^{\circ}$
using slit widths ${\Delta}y = $ 5, 7, and 10 pixels (14, 20, and 28 Mm, respectively).  These three
panels show the same collection of quasi-parallel tracks that are slanted from the lower left to
the upper right.  The  tabulated speeds of 0.30${\pm}$0.02 km s$^{-1}$ refer to the slant near the
mid-point of the chord and the accuracy of my eye-estimates of their slopes. In addition, the tracks
bend horizontally toward the east and west limbs (at the bottom and top of the maps, respectively).
This is partly due to the curvature of the Sun's surface
which causes the speed to change from a sky plane motion at the central meridian to a more
line-of-sight motion toward the limbs.  If the $B_{0}$ angle were 0, the latitude contours would
be straight lines along the slit.  In that case, the speed, $v$ and $x$-position could be described
in terms of a longitudinal angle, ${\phi}$, using the equations:
\begin{subequations}
\begin{align}
v~=~v_{1}\cos{\phi}\\
x~=~R_{\odot}\cos{\lambda}_{s}\sin{\phi},
\end{align}
\end{subequations}
where $v_{1}$ is the surface speed, ${\lambda}_{s}$ is south latitude, and $R_{\odot}$ is
the solar radius.  Eliminating ${\phi}$ between these two equations, we are left with a relation
between the speed and the distance along the slit.
\begin{equation}
\frac{v}{v_{1}}~=~\sqrt{1-(\frac{x}{R_{\odot}\cos{\lambda}_{s}})^2}.
\end{equation}
\begin{figure}[t!]
%\plotone{fig02.pdf}
 \centerline{
 \fbox{\includegraphics[bb=22 15 592 760,clip,width=0.83\textwidth]
 {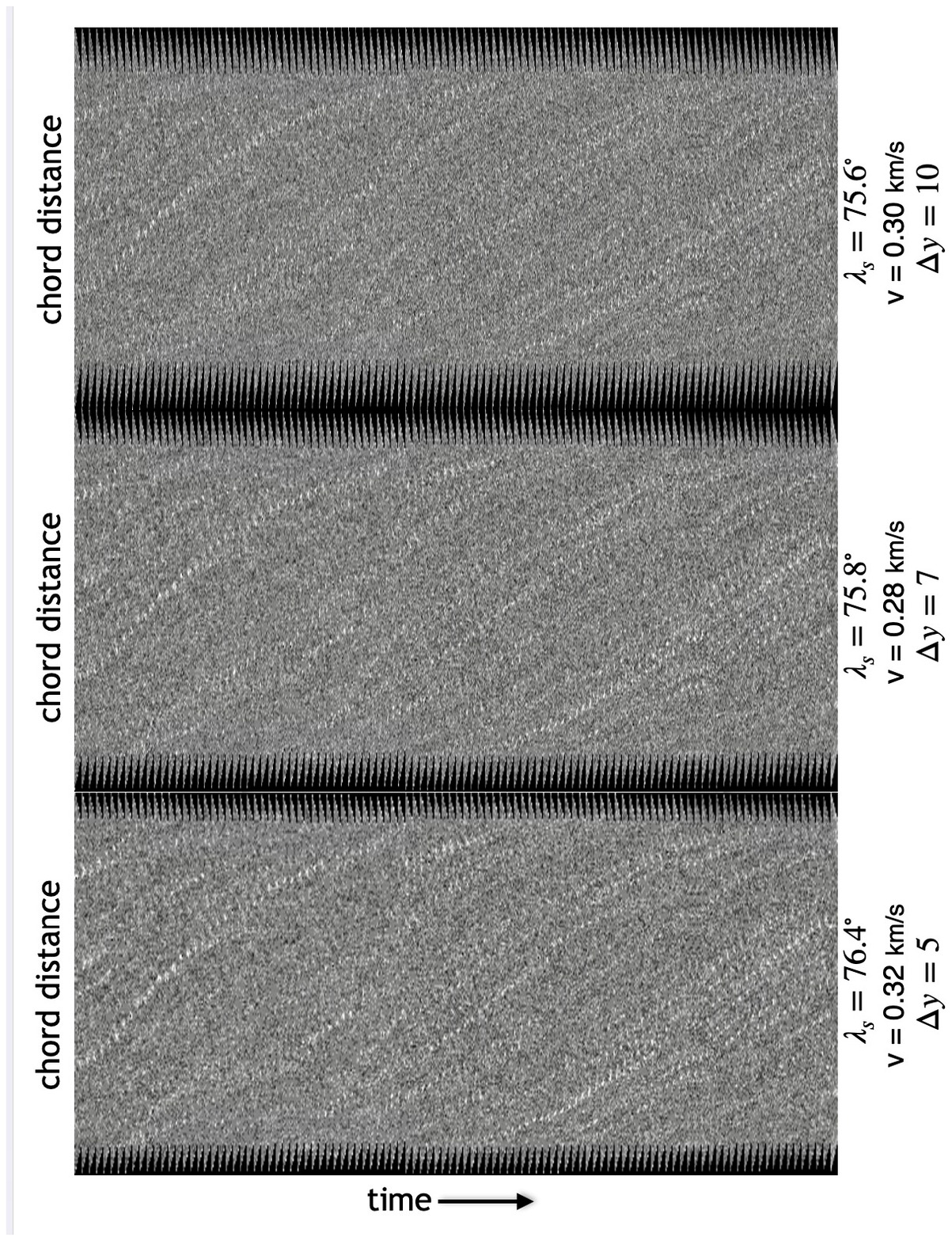}}}
\caption{Space-time maps obtained at approximately 76$^{\circ}$ latitude, with slit widths,
${\Delta}y = $ 5, 7, and 10 pixels (14, 20, and 28 Mm, respectively), showing a greater
latitudinal compression and lower resulting contrast of faculae observed with the wider slits.  The
chord distance is approximately $2R_{\odot} \cos({\lambda}_{s}-6.8^{\circ})$ and the time
duration is 30 days.  The associated speeds are based on eye-estimates of the average inclinations,
and their ${\pm}$ 0.02 km s$^{-1}$ differences reflect the accuracy of those estimates.}
\label{fig:fig1}
\end{figure}
\pagebreak

Expressing $x$ and $v$ in dimensionless units, and recognizing that $v = dx/dt$, this equation
becomes
\begin{equation}
\frac{dx}{dt}~=~\sqrt{1-x^2},
\end{equation}
whose solution is simply $x = \sin{t}$.  Restoring the units, this solution is
\begin{equation}
\frac{x}{R_{\odot}\cos{\lambda}_{s}}~=~\sin \left ( \frac{v_{1}t}{R_{\odot}\cos{\lambda}_{s}} \right ).
\end{equation}
So, the tracks are simple sine waves whose arguments run from -${\pi}/2$ to +${\pi}/2$.
For $v_{1}t/R_{\odot}\cos{\lambda}_{s} << {\pi/2}$, the solution is $x=v_{1}t$, and in a space-time map, the tracks would have a constant slope, $v_{1}$, around the mid-point of 
the chord.  However, as the time, $t$, increases, and the quantity,  $v_{1}t/R_{\odot}\cos{\lambda}_{s}$
approaches ${\pi}/2$, the line bends horizontally toward the upper end of the chord where
$x=R_{\odot}\cos{\lambda}_{s}$ and $dx/dt = 0$.   Of course, the antisymmetric behavior occurs in the
eastern hemisphere where $x$ is negative and the curve approaches the lower end of the chord.
As mentioned in the previous section, when $B_{0}=-6.8^{\circ}$, I ignore the mismatch between the straight slit and the curved latitude profiles, and determine the central speed, $v_{1}$, by extrapolating the linear segment of the track to the top and bottom of the space-time  map and divide the chord length by the duration of time, ${\Delta}t$, that is intercepted between the ends of the slit.
\begin{figure}[h!]
%\plotone{fig03.pdf}
 \centerline{
 \fbox{\includegraphics[bb=53 15 560 775,clip,width=0.85\textwidth]
 {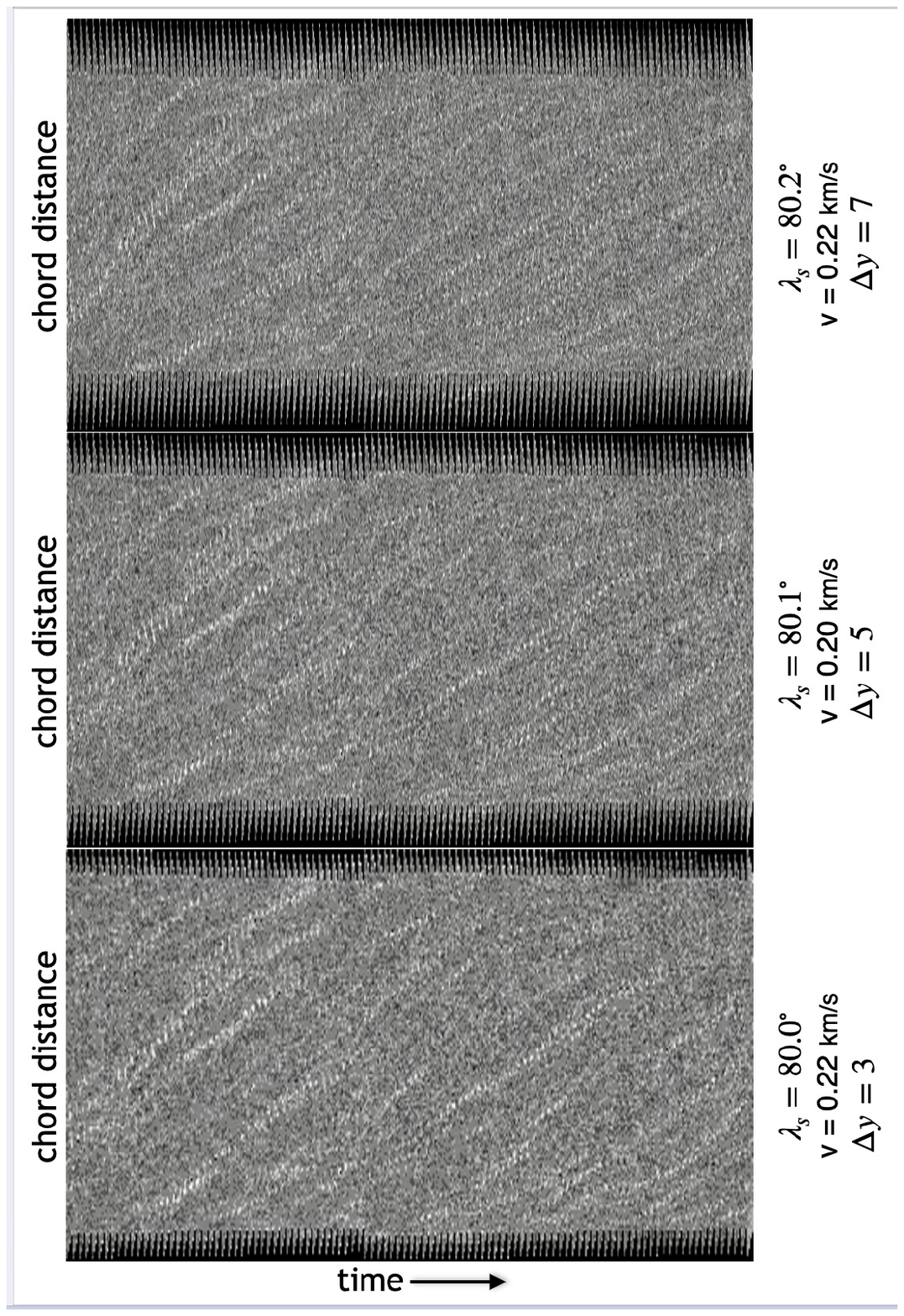}}}
\caption{Same as Figure~2, except for slit widths, ${\Delta}y = $ 3-, 5-, and 7-pixels
(8, 14, and 20 Mm, respectively) and a common latitude ${\lambda}_{s} {\sim}$80$^{\circ}$,
again showing the greater latitudinal compression and decreased contrast obtained with
the wider slits.}
\label{fig:fig1}
\end{figure}

\begin{figure}[h!]
%\plotone{fig04.pdf}
 \centerline{
 \fbox{\includegraphics[bb=20 15 590 770,clip,width=0.95\textwidth]
 {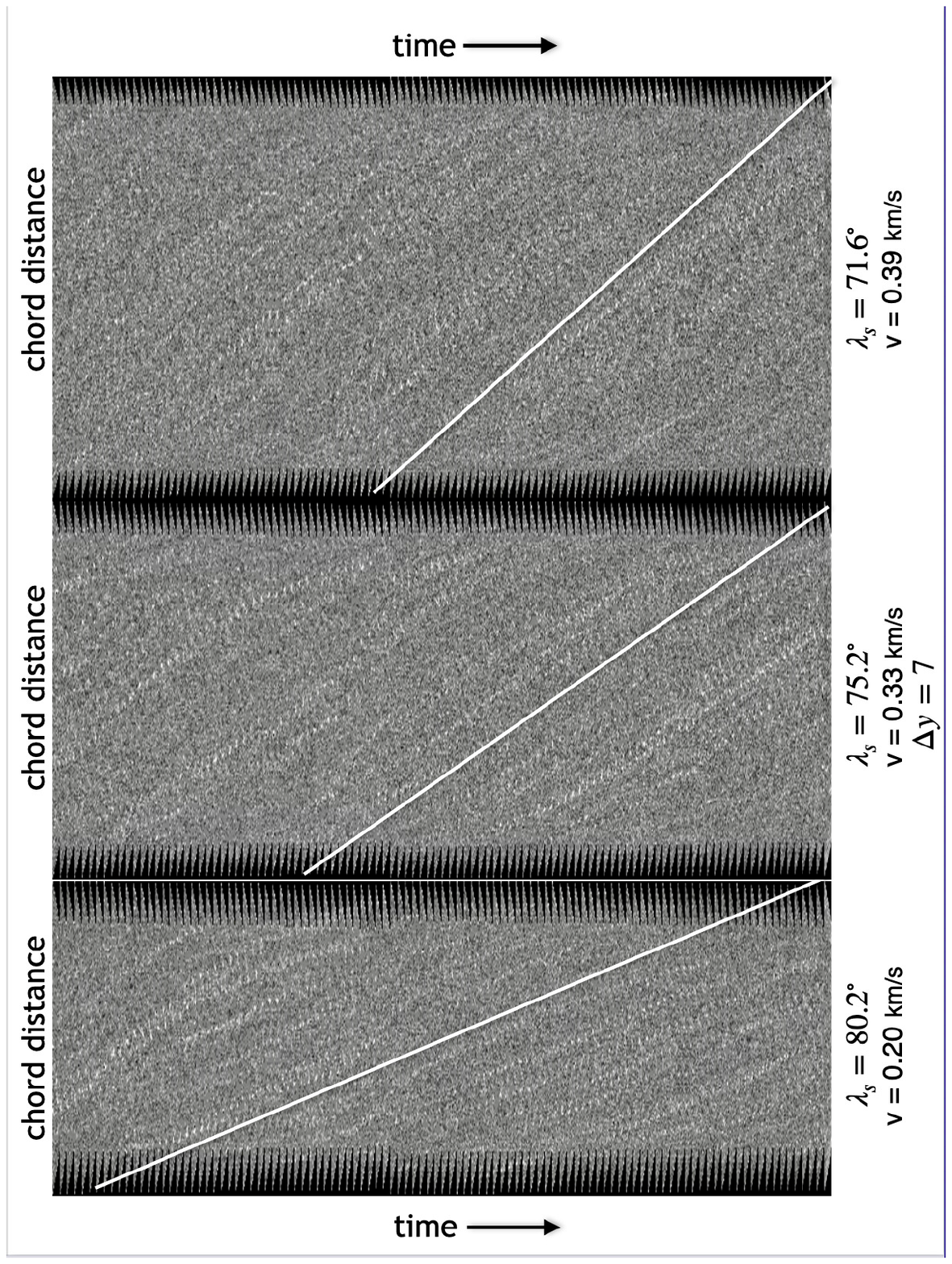}}}
\caption{Space-time maps obtained using a 7-pixel wide slit (20 Mm) at chords located
at ${\lambda}_{s} = $80.2$^{\circ}$, 75.2$^{\circ}$, and 71.6$^{\circ}$ south latitude, showing
tracks of  polar faculae during 7-21 February 1997 and 1998.  The lengths of the
vertical and horizontal axes are $\sim$2R$_{\odot}\cos({\lambda}_{s}-6.8^{\circ})$ and 30 days,
respectively.  White lines are guides for identifying the average slopes of the tracks.}
\label{fig:fig1}
\end{figure}
\noindent

Another important characteristic of the space-time maps in Figure~2 is their slit-dependent
differences.  In order to display these panels with the same 30-day horizontal dimension, it was
necessary to compress the 10-pixel slits twice as much as the 5-pixel slits.  Likewise, the
7-pixel slits were compressed 1.4 times as much as the 5-pixel slits.  This horizontal
matching procedure caused the faculae to be narrower in the wider-slit images than in the
narrower-slit images.  With careful scrutiny, one can see the individual faculae changing from
fine, elongated features in the 10-pixel map to rounder features in the 5-pixel map.  This
compression seems to reduce the contrast of the faculae so that the tracks are visible with greater contrast in the space-time maps made with the narrower slits. 

In each map, the time axis spans 30 days with 106 strips whose jagged edges are clearly visible at
the top and bottom of the map.  As mentioned in the previous subsection, the cadence of these
strips is uniform at the rate  of 4 strips per day in the second half of the map (1998) and variable
at an average rate of about 3 strips per day in the first half of the map (1997).  Despite this variable cadence, most of the tracks seem fairly well defined, with steep slopes near the central meridian
(the half-chord location) and lower slopes toward the east and west limbs (at the bottom and top
of each map).   

Figure~3 is a similar comparison using smaller slit widths of ${\Delta}y = $ 3, 5, and 7 pixels and a
larger average latitude at 80$^{\circ}$ south latitude.  The relatively high latitude and narrow slit
widths were chosen to increase the visibility of the tracks.  In these
80$^{\circ}$ maps, the tracks seem to have higher contrast than they did at the lower latitude,
even when comparing maps of the same slit width.  Each panel shows virtually identical
patterns of quasi-parallel tracks, moderately steep around the central meridian and shallow toward
the limbs.  In addition, the tracks seem more fragmentary than in the lower-latitude space-time
maps of Figure~2, especially toward the limbs.  Presumably, this is a consequence of a greater
mismatch between the straight slit and the curved latitude contours at the higher latitude.

Nevertheless, it is still possible to measure their average slope.  An eye-estimate
of the average mid-latitude slope gives the speeds of 0.20-0.22 km s$^{-1}$ shown in the
figure, where again the differences reflect the accuracy of those estimates.  Clearly, this
accuracy is sufficient to show the decrease of speed from 0.3 km s$^{-1}$ to 0.2 km s$^{-1}$
as the latitude increases from 76$^{\circ}$ to 80$^{\circ}$.
 
Figure~4 supports this trend by including a third latitude, and provides straight lines as guides
for identifying the average slopes.  With these space-time maps, the measurement indicates
the average contribution of all the faculae and does not depend on tracking a single facula during its relatively short $\sim$1-day lifetime.  (In fact, the relatively short lifetimes of individual polar faculae
were probably the main reason that 2005 summer student, Hanna Krug, and I had difficulty measuring
the high-latitude rotation profile by tracking individual polar faculae on these same SOHO/MDI solar images.)  Not only do the white lines decrease their inclinations from the upper panel to the lower
panel, but the corresponding speeds decrease from 0.39 km s$^{-1}$ at ${\lambda}_{s}=71.6^{\circ}$
to 0.33 km s$^{-1}$ at ${\lambda}_{s}=75.2^{\circ}$, and finally to 0.20 km s$^{-1}$ at
${\lambda}_{s}=80.2^{\circ}$.  Next we plot the results for all of our measurements.
\pagebreak
 
\section{Results}  
Figure~5 shows a plot of tracking measurements from space-time maps like those in
Figures~2-4, except that the measurements were obtained using a relatively wide
10-pixel slit, corresponding to about 28 Mm.  In each case, the latitude refers to the average
latitude of the upper and lower edge of the slit.  The data points were fit differently in
the left and right panels of Figure~5.  In the left panel, the red dashed line indicates
the usual 2-parameter best fit, and leads to the values of the slope (1.54 km s$^{-1}$)
and polar intercept (0.082 km s$^{-1}$), as indicated by the formula
\begin{equation}
\frac{(v-0.082)}{1.54}~=~-\frac{({\lambda}_{s}-90)}{90},
\end{equation} 
where ${\lambda}_{s}$ is the south latitude in degrees; the velocity, $v$, and fit parameters
are in km s$^{-1}$.  Also, the rms deviation of this line from the measured speeds is given
as 0.0180 km s$^{-1}$.  So the fit is fairly good, but the line does not pass through the
point at (90,0), indicated by the small black circle.
\begin{figure}[h!]
%\plotone{fig05.pdf}
 \centerline{
 \fbox{\includegraphics[bb=110 280 500 685,clip,width=0.50\textwidth]
 {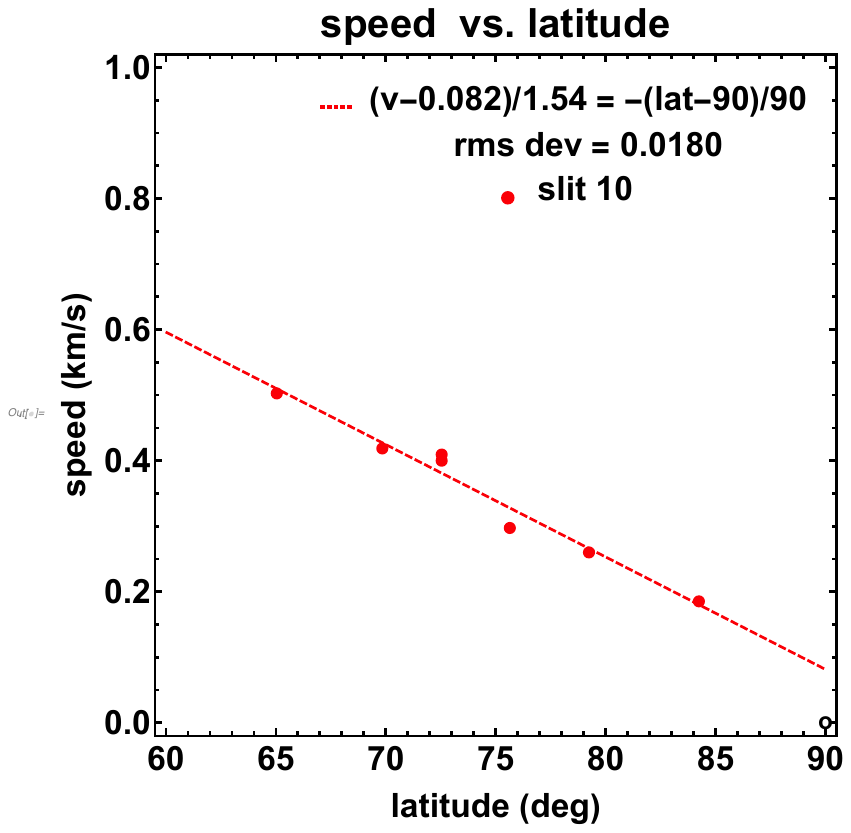}}
 \hspace{0.01in} 
  \fbox{\includegraphics[bb=110 280 500 685,clip,width=0.50\textwidth]
 {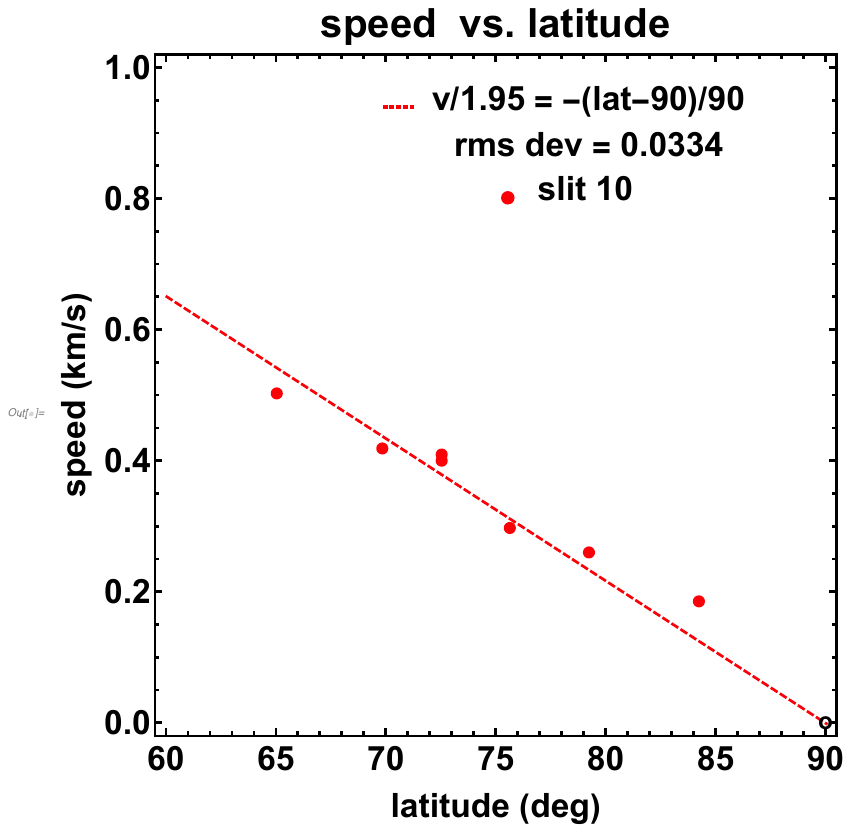}}}
\caption{Speed versus latitude for measurements obtained with a 10-pixel slit (28 Mm).
(left): a 2-parameter straight-line fit for slope and polar speed;
(right): a 1-parameter fit for the slope of a line passing through the point (90,0).  The rms
deviations are ${\sim}$0.02 km s$^{-1}$ and ${\sim}$0.03 km s$^{-1}$, respectively, but
the slope is steeper when the speed is forced to vanish at the pole.}
\label{fig:fig1}
\end{figure}

The terminology is the same in the right panel, but the straight line is determined differently.
The line is forced to pass through the black circle at the point (90,0), and the slope of the line is
taken as the single, best-fit parameter.  The result is
\begin{equation}
\frac{v}{1.95}~=~-\frac{({\lambda}_{s}-90)}{90},
\end{equation}
and the amount of scatter is 0.0334 km s$^{-1}$, slightly larger than 0.0180 km s$^{-1}$
obtained with the 2-parameter fit in the left panel.  An important contribution of this
forced polar fit is to increase the value of the speed parameter from 1.54 km s$^{-1}$
to 1.95 km s$^{-1}$.  This has a significant effect on the resulting synodic rotation speed,
as we will see below.

Although this result with the 10-pixel slit seems pretty good, it is not as good as I obtained
using smaller slits.  Figure~6 shows a similar result using slit widths of 7, 5, and 3 pixels
($\sim$20 Mm, 14 Mm, and 8 Mm, respectively).  These measurements are indicated by red
points, blue points, and a single green point for the 3-pixel slit.  In this case, the left and right
panels give 
comparable amounts of scatter (0.023 km s$^{-1}$ compared to 0.025 km s$^{-1}$), with
a `polar
\begin{figure}[h!]
%\plotone{fig06.pdf}
 \centerline{
 \fbox{\includegraphics[bb=110 280 500 685,clip,width=0.50\textwidth]
 {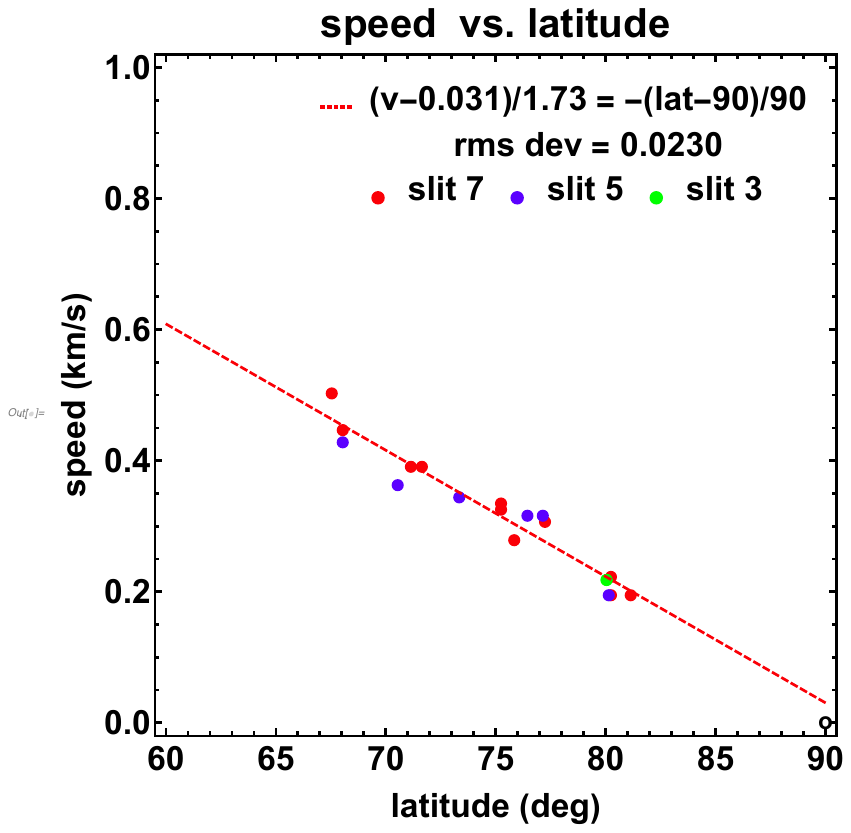}}
 \hspace{0.01in} 
  \fbox{\includegraphics[bb=110 280 500 685,clip,width=0.50\textwidth]
 {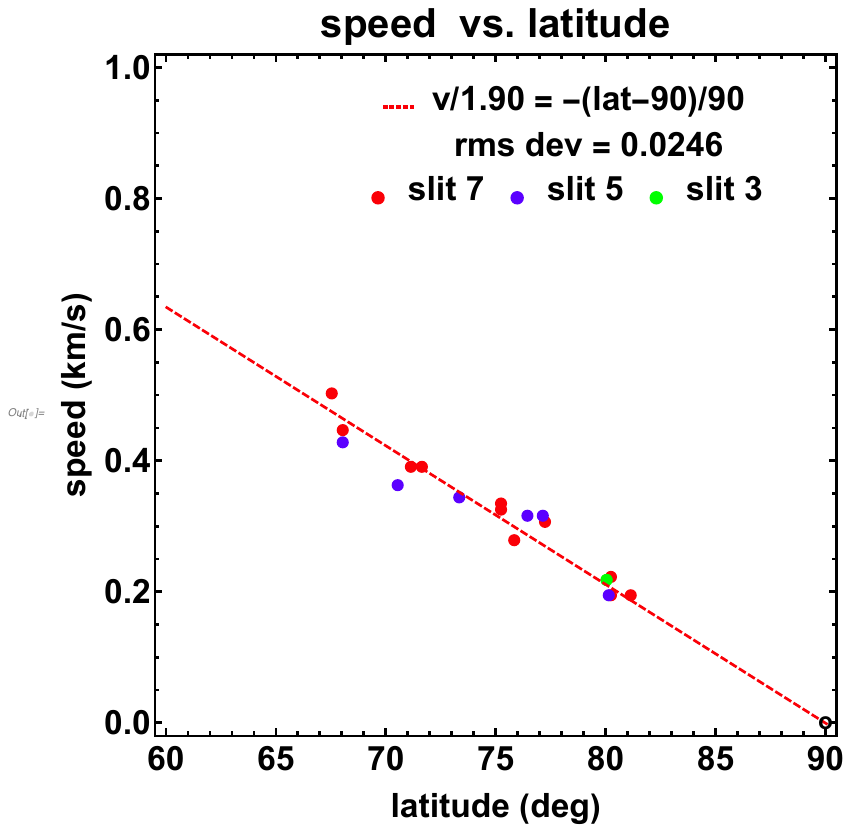}}}
\caption{Same as Figure~5, but with narrower slits (7-, 5-, and 3-pixel widths are indicated
by red, blue, and green points, respectively).   With these narrower slits, the 2-parameter fit
projects to 0.03 km s$^{-1}$ at the pole with an rms deviation of 0.023 km s$^{-1}$, which is
slightly smaller than the 0.025 km s$^{-1}$ deviation obtained when the speed is forced to
vanish at the pole.}
\label{fig:fig1}
\end{figure}
\noindent
 intercept' of 0.03 km s$^{-1}$.  When the speed is forced to vanish at the pole,
the speed parameter increases from 1.73 km s$^{-1}$ to 1.90 km s$^{-1}$ as shown here:
\begin{equation}
\frac{v}{1.90}~=~-\frac{({\lambda}_{s}-90)}{90}.
\end{equation}
\begin{figure}[h!]
%\plotone{fig07.pdf}
 \centerline{
 \fbox{\includegraphics[bb=110 280 500 685,clip,width=0.50\textwidth]
 {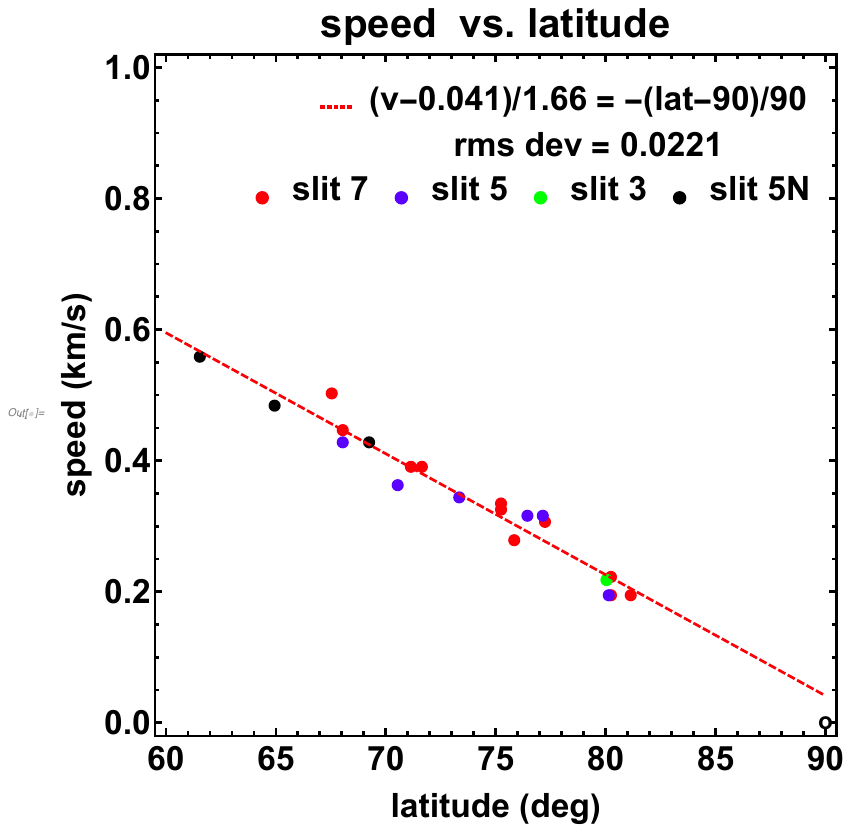}}
 \hspace{0.01in} 
  \fbox{\includegraphics[bb=110 280 500 685,clip,width=0.50\textwidth]
 {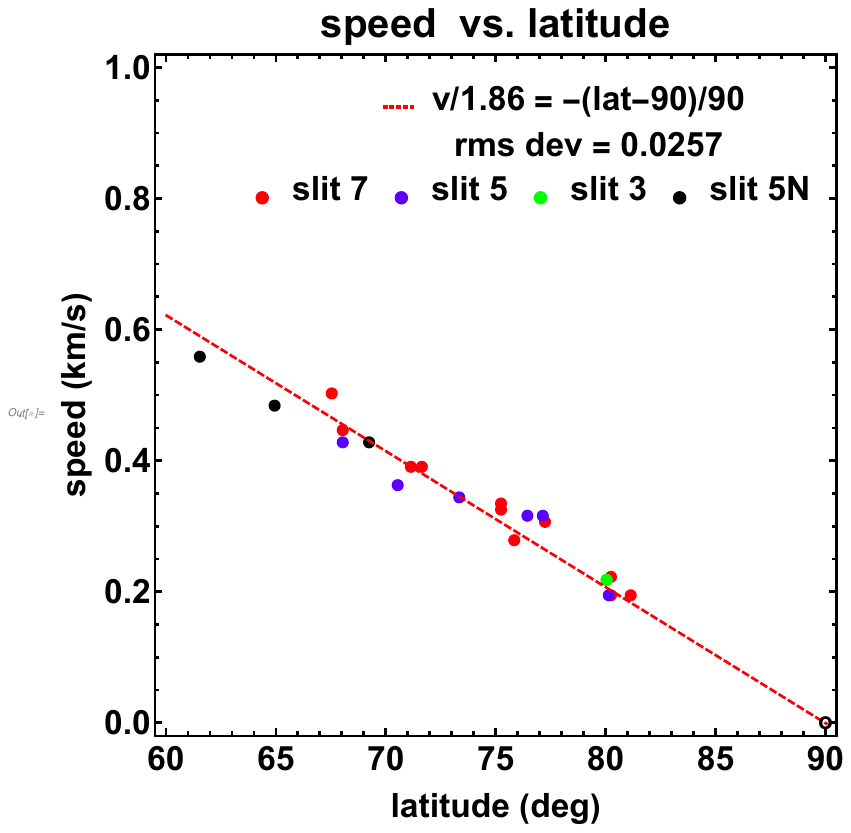}}}
\caption{Same as Figure~6, but with the addition of 3 measurements of north polar faculae
in the 60-70$^{\circ}$ latitude range with a 5-pixel slit.  The resulting straight-line fit is
slightly better than the fit obtained without those extra points, and the forced fit has a slightly
smaller speed parameter (1.86 km s$^{-1}$ compared to 1.90 km s$^{-1}$).}
\label{fig:fig1}
\end{figure}

Finally, Figure~7 shows the same data with the addition of 3 points obtained in the north polar cap
with a 5-pixel slit.  These northern-hemisphere points lie in range of 60-70$^{\circ}$, which
is systematically lower than the range obtained for the southern-hemisphere points due to the
unfavorable view of the north polar region during 7-21 February.   In practice, when I calculated
the linear offset from the polar limb, I obtained $\sin({\lambda}_{s}-6.8)$ in the south
and  $\sin({\lambda}_{n}+6.8)$ in the north.  Thus, for the same offset, it was necessary to add
 6.8$^{\circ}$ in the south and subtract 6.8$^{\circ}$ in the north, making the northern-hemisphere
latitude range about 13.6$^{\circ}$ lower than the southern-hemisphere range.  This had the effect
of extending the latitude range of the measurements to slightly lower latitudes and slightly reducing
the 2-parameter rms scatter from 0.023 to 0.022 km s$^{-1}$.  Also, the speed parameter of the
 forced fit decreased slightly from 1.90 to 1.86 km s$^{-1}$.  This was a relatively minor
 change;  probably the most that we can conclude from this comparison is that the
 north polar-cap measurements are consistent with the south polar-cap measurements,
as one might expect if the two polar regions have the same rotation rate.

Based on these measurements, we conclude that the high-latitude linear speed, $v$, is given
by an expression of the form
\begin{equation}
\frac{v({\lambda}_{s})}{v_{0}}~=~-\frac{({\lambda}_{s}-{\pi}/2)}{{\pi/2}},
\end{equation}
where $v_{0}$ ${\approx}$ 1.90${\pm}$0.05 km s$^{-1}$, and ${\lambda}_{s}$ is south latitude, expressed
in radians.  As I mentioned in the Introduction, the significance of this relation is that
$v({\lambda}_{s})$ and $\cos{\lambda}_{s}$ have the same ${\lambda}_{s}$-dependence
close to the south pole.  This means that their ratio, and thus the angular rotation rate, ${\omega}$,
is constant close to the pole.  In particular,
\begin{equation}
{\omega}~=~\frac{v({\lambda}_{s})}{R_{\odot}\cos{\lambda}_{s}}~=~
\frac{v_{0}(2/{\pi})({\pi}/2-{\lambda}_{s})}{R_{\odot}\sin({\pi}/2-{\lambda}_{s})}~
{\approx}~\frac{2}{{\pi}}\frac{v_{0}}{R_{\odot}},
\end{equation}
when $({\pi}/2-{\lambda}_{s}) << 1.$
Substituting $v_{0}$ = 1.90 km s$^{-1}$ and R$_{\odot}$ ${\approx}$ 0.7 ${\times}$10$^{6}$ km,
and converting from radians to degrees, I get ${\omega}~{\approx}~$ 8.6$^{\circ}$ day$^{-1}$
as the synodic rotation rate of the south polar cap.  The sidereal rate would be about
1$^{\circ}$ day$^{-1}$ faster.  (We can be sure that Eq(8) refers to the synodic rate because it
was obtained by tracking solar features.  The procedure is analogous to placing a mark on the Sun with a grease pencil and recording the elapsed time until the Sun's rotation returns that mark to the central meridian as seen from the solar orbiting Earth and SOHO spacecraft.)  

\section{Summary and Discussion}
It is easy to summarize this work. With the help of ChatGTP, I made space-time maps from
SOHO/MDI 6767 {\AA} continuum images of the Sun's south pole in 7-21 February 1997 and 1998.
Those maps showed many parallel tracks produced by the solar rotation of polar faculae.  By
measuring the slopes of those tracks, I obtained speeds whose values decreased with latitude in
the 60-80$^{\circ}$ range and projected linearly to 0 km s$^{-1}$ at the south pole.  This gave a constant synodic rotational speed of about 8.6 $^{\circ}$ day$^{-1}$ in a small polar cap at the Sun's south pole, as one would expect for a polar asymptote of the Sun's rotation profile
\citep{SDeV_1986}.  A few measurements of north polar faculae gave values consistent
with those obtained for south polar faculae, supporting the idea that the two polar regions
rotate at the same rate.

The next question is how do these measurements compare with past measurements of the
Sun's high-latitude rotation.   Figure~8 summarizes some of the earlier measurements.  The blue
dashed curve shows the polar extrapolation of the well-known \cite{SNOD_1983} rotation
profile, and the dashed red curve shows the results of this paper as summarized
\begin{figure}[h!]
%\plotone{fig08.pdf}
 \centerline{
 \fbox{\includegraphics[bb=110 280 500 685,clip,width=0.70\textwidth]
 {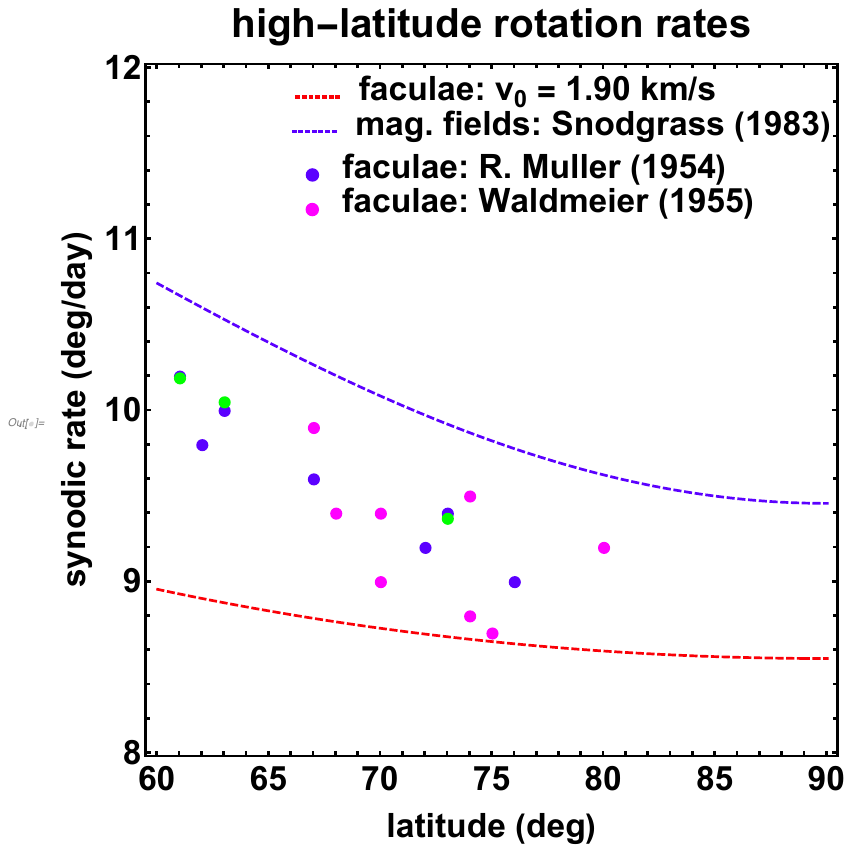}}}
\caption{A comparison of the synodic rotation rates obtained in this paper with high-latitude
measurements obtained previously.  The dashed blue curve is a best fit extrapolation of
the \cite{SNOD_1983} tracks of magnetic flux elements, and the dashed red curve refers
to the result given in Eqs (8) and (9) of this paper.  Previous measurements of polar faculae
by \cite{ROLF_1954} and \cite{WALD_1955} are shown in the region between those two curves. }
\label{fig:fig1}
\end{figure}
\noindent
by Eqs (8) and (9) with $v_{0}=1.90$ km s$^{-1}$.
At the pole, the two curves differ by about 0.9 $^{\circ}$ day$^{-1}$, corresponding to synodic
rotation periods of about 38 days (Snodgrass) and 42 days (this paper).

The colored dots between
these two curves indicate polar faculae measurements by \cite{ROLF_1954} and \cite{WALD_1955}.
The blue dots of Waldmeier are reasonably well mixed with the pink dots of Muller.  The green dots
that are nearly coincident with blue dots are Muller's `best measurements', obtained for especially
long-lived faculae.  In his paper, \cite{WALD_1955} mentions that the point at 80$^{\circ}$ latitude
is `significantly less precise than the others due to its high latitude'.  By contrast, he says that
`point 6' at (9.5$^{\circ}$ day$^{-1}$, 74$^{\circ}$ latitude) `appears to be particularly reliable'.
It is interesting that Muller's 3 best measurements and Waldmeier's particularly reliable measurement
are consistent with each other and with Waldmeier's significantly less precise measurement; also,
those `most reliable' measurements lie roughly alongside the Snodgrass curve with a slope that
is steeper than the more distant  red curve that I obtained by tracking polar faculae in 1997-1998.

In Figure~8, both sets of data points were obtained in the early 1950s when polar faculae occurred in greater numbers than at any time during the past 100 years \citep{SHEfac_2008}.  Mueller observed
them between September 1952 and April 1954, and Waldmeier observed them during 1951-1954.  By comparison, the measurements described in this paper were obtained in 1997 and 1998 at the end
of the last cycle of moderately strong polar fields and modest numbers of polar faculae.  Probably the
opportunity for such a study would have been less favorable during the sunspot minima of 2009
and 2020 when the polar fields were relatively weak.  Similarly, there may have been a reduction in the numbers of polar faculae in the weak sunspot cycles during 1880-1910, which might explain why the Greenwich observers did not find a relation between the sunspot cycle and the numbers of polar
faculae in those years \citep{GREN_1923,KIEP_1953}.

Despite the good agreement between the high-latitude measurements of \cite{ROLF_1954} and
 \cite{WALD_1955}, those rotation rates do not provide a good match with either the speeds
 of magnetic tracers observed by \cite{SNOD_1983} or with the measurements of this paper.
 One might understand the disagreement with the magnetic flux elements because they were
 observed with relatively low spatial resolution using the Mount Wilson Observatory (MWO)
 magnetograph.  On the other hand, the images from the
 SOHO/MDI instrument were free of ground-based seeing conditiions and probably had a
 spatial resolution that was consistently better than that obtained by the 1950s-era ground-based telescopes, especially as the seeing degraded while the faculae were being tracked during each day.  Also, in the 1950's, neither Muller nor Waldmeier had digital data that could be processed by
 computers and analyzed with artificial intelligence like ChatGPT.  But those early observers
were known for the high quality of their work, and it seems likely that their polar faculae measurements are fairly accurate.  Considering that I used a new and untested technique, it seems
more likely that I could have overlooked something that may have caused the discrepancy.
 
 Consequently, I have reexamined several aspects of my measurement procedure.  It seemed
 to make more sense to calculate the linear speeds first, finding a best-fit solution and then
 converting from linear to angular speed by dividing the solution by $R_{\odot}\cos{\lambda}_{s}$,
 the radius of the latitude contours.  This would avoid the large amount of scatter that would occur
 if the measured speeds were divided by a progressively smaller number as $\cos{\lambda}_{s}$
 approached 0 at the pole.  A linear speed profile that reached 0 km s$^{-1}$ at the pole
 made sense because it offset this $\cos{\lambda}_{s}$ behavior and gave a polar asymptote
 similar to that found at the equator.  So, if there were an error, I would expect it to be a systematic
 error, perhaps produced in the calibration of the chord distance or the elapsed time.  Such a
 systematic error would change the speeds uniformly without affecting their approach to the pole.
 
 For example, the synoptic rotation rate is very sensitive to the effective latitude used in the
 the plots of speed versus latitude.  If the average latitude, $<{\lambda}_{s}>$, that I used for
 the plots were 1$^{\circ}$ too small, it would cause the data points in Figures~5-7 to be
 1$^{\circ}$ too far to the left.  Consequently, a correction would shift the points to the right
 so that a forced fit to those corrected points would increase the value of $v_{0}$ to about
 2.0 km s$^{-1}$.  This would increase the synodic rotation rate, ${\omega}$, from
 8.6$^{\circ}$ day$^{-1}$ to 9.0$^{\circ}$ day$^{-1}$, and place my dashed red curve in the
 midst of the polar faculae measurements (but with a lower slope than one would infer from those measurements).
 
 There are several ways that these measurements could be improved.  First, the
 space-time slit could be curved so that it fits the shape of the polar latitudes.  This might
 strengthen the tracks, and also allow measurements closer to the pole.  Second, one might try
 using continuum images from the \textit{Solar Dynamics Observatory} (SDO) spacecraft.
 Presumably, these images were obtained uniformly and at a greater cadence than those
 obtained with the SOHO/MDI instrument.  Also, one could study the effect of $B_{0}$ by
 selecting times that it is 7.25$^{\circ}$ (for tracking north polar faculae), -7.25$^{\circ}$ (for
 tracking south polar faculae), and 0$^{\circ}$ (when the latitude contours would be straight
 and well suited for a straight space-time slit.)  Possible disadvantages might be a limited
 number of polar faculae in recent years when the polar fields have been relatively weak,
 and the need to flat-field the solar images. (Flat fielding might be relativley easy with the help of
 ChatGPT.)  (Since preparing this paper, I learned that Pesnell, Clark, and Barzal made a
 presentation at the COFFIES workshop, in which they showed a flat-fielded image obtained
 in the 4500 {\AA} continuum with the Atmospheric Imaging Assembly (AIA) on SDO.  The image
 was created using a `progressive standard deviation method', and showed horizontal lines
(or arcs) indicating the positions of polar faculae during the course of a day.)  Third, it would be interesting to try using very high spatial resolution like that available with the 4 m Daniel K. Inouye
Solar Telescope (DKIST) on Maui, Hawaii or the 1.6 m Goode Solar Telescope at Big Bear Solar Observatory.  Such high resolution might allow measurements very close to the pole (or on
the other side of the pole that is favorably tipped toward the Earth).

%\begin{acknowledgments}
I am grateful to Yi-Ming Wang (NRL) for telling me about the presentations at the
26 June 2024 COFFIES workshop on the `Science at the Poles'.  I am also grateful to
Harry Warren (NRL) for helping me to analyze the 6767 {\AA} continuum images that were
obtained by the SOHO/MDI instrument during 1996-2005 and that led to our original paper about the distribution of faculae on the Sun \citep{SHEWAR_2006}.
%\end{acknowledgments}


%\vspace{5mm}
%\facilities{HST(STIS), Swift(XRT and UVOT), AAVSO, CTIO:1.3m,


%\software{astropy \citep{2013A&A...558A..33A,2018AJ....156..123A},  
   %       Cloudy \citep{2013RMxAA..49..137F}, 
 %         Source Extractor \citep{1996A&AS..117..393B}
  %        }

%\appendix

%\section{Ylm calculations}
%Here are some details.

%\section{I, J, K and the effect of B0}
%Here are more details.


\bibliography{ms}{}
\bibliographystyle{aasjournal}
%\bibliographystyle{apj}



\end{document}